%% file: main.tex
\documentclass[journal=jacsat,manuscript=article]{achemso}
\usepackage[utf8]{inputenc}
\usepackage[T1]{fontenc}

\usepackage[version=3]{mhchem} 
\usepackage[usenames,dvipsnames]{xcolor}
\usepackage{lipsum}
\usepackage[normalem]{ulem}

\newcounter{Paragraph}
\author{Yongkang Wang}
\affiliation{Max
Planck Institute for Polymer Research, Ackermannweg 10, 55128 Mainz,
Germany}
\alsoaffiliation{%
These authors contributed equally}
\email{wangy3@mpip-mainz.mpg.de}

\author{Yair Litman}
\affiliation{Max
Planck Institute for Polymer Research, Ackermannweg 10, 55128 Mainz,
Germany}
\alsoaffiliation{Yusuf Hamied Department of Chemistry,  University of Cambridge,  Lensfield Road,  Cambridge,  CB2 1EW, UK}
\alsoaffiliation{%
These authors contributed equally}
\email{litmany@mpip-mainz.mpg.de}

\author{Minhaeng Cho}
\affiliation{Center for Molecular Spectroscopy and
Dynamics, Institute for Basic Science (IBS), Seoul 02841,
Republic of Korea; Department of Chemistry, Korea
University, Seoul 02841, Republic of Korea}

\author{Stephen Cox}
\affiliation{Department of Chemistry, Durham University, South Road, Durham, DH1 3LE, United Kingdom}

\author{Mischa Bonn}
\affiliation{Max
Planck Institute for Polymer Research, Ackermannweg 10, 55128 Mainz,
Germany}
\email{bonn@mpip-mainz.mpg.de}

\title[An \textsf{achemso} demo]{
Wetting Transparency of Graphene: A macroscopic Window but Nanoscopic Mirror 
}

\begin{document}

\begin{tocentry}




\vspace{0mm}
\includegraphics[width=0.95\textwidth]{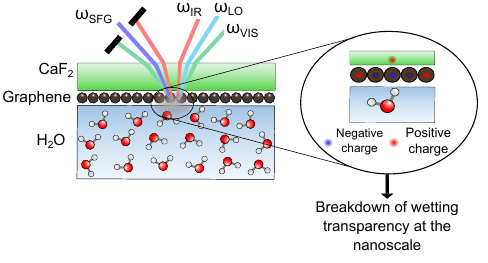}


\end{tocentry}


\input{abstract}

\input{body}

~
~

The authors declare no competing financial interest.

\begin{acknowledgement}

We are grateful for the financial support from the MaxWater Initiative of the Max Planck Society. Funded by the European Union (ERC, n-AQUA, 101071937). Views and opinions expressed are however those of the author(s) only and do not necessarily reflect those of the European Union or the European Research Council Executive Agency. Neither the European Union nor the granting authority can be held responsible for them. S.J.C is a Royal Society University Research Fellow at Durham University (URF\textbackslash R1\textbackslash 211144). M.C. is grateful for financial support from the Institute for Basic Science in Korea (IBS-R023-D1).

\end{acknowledgement}

\begin{suppinfo}

Graphical representation of the computational setup, density profiles of ions,  further water orientation 2D profiles, and additional details on the experimental data analysis can be found in the SI.

\end{suppinfo}


\providecommand{\latin}[1]{#1}
\makeatletter
\providecommand{\doi}
  {\begingroup\let\do\@makeother\dospecials
  \catcode`\{=1 \catcode`\}=2 \doi@aux}
\providecommand{\doi@aux}[1]{\endgroup\texttt{#1}}
\makeatother
\providecommand*\mcitethebibliography{\thebibliography}
\csname @ifundefined\endcsname{endmcitethebibliography}
  {\let\endmcitethebibliography\endthebibliography}{}

\end{document}


\clearpage

\begin{figure}
\includegraphics[width=0.6\textwidth]{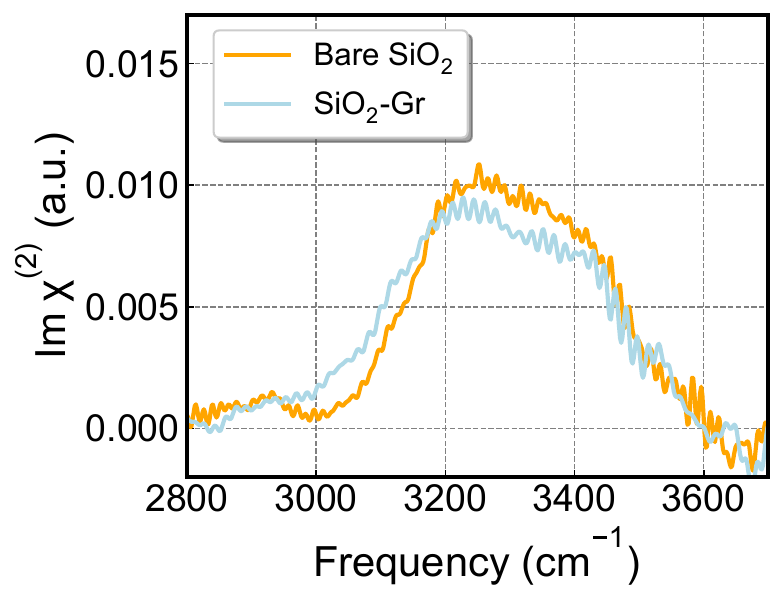}
  \caption{\textbf{Graphene has little effect on the SiO$_2$/water interaction near neutral pH} The O-H stretching Im$\chi^{(2)}$ spectra at the bare SiO$_2$/water and  SiO$_2$-supported graphene/water interfaces measured with 10 mM NaCl at pH 6.}
  \label{fig:Sio2}
\end{figure}

Figure~\ref{fig:Sio2} shows the Im$\chi^{(2)}$ spectra of bare SiO$_2$/water and SiO$_2$-supported graphene/water interfaces. As for the CaF$_2$ substrate, graphene induces only minor changes in the interfacial water structure.

\begin{figure*}
\includegraphics[width=.45\columnwidth]{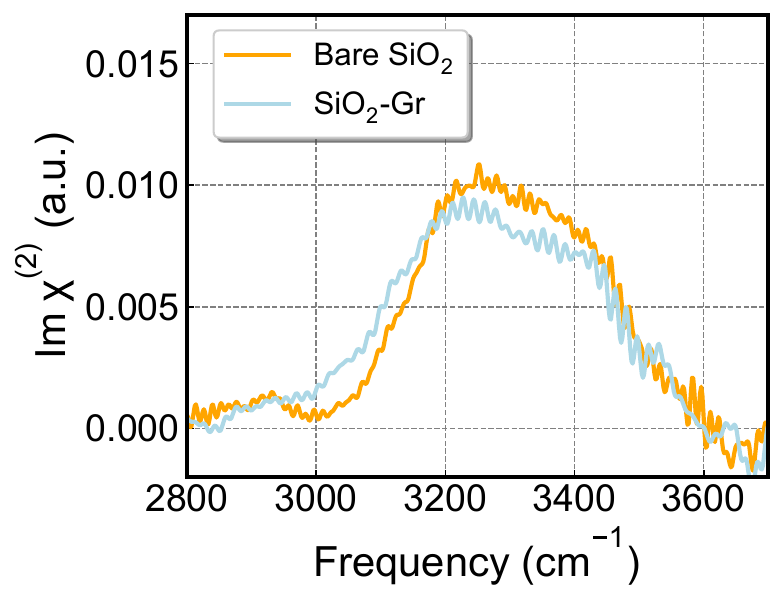}
  \caption{\textbf{Graphene-induced polarization reshapes the initial interfacial water layer but vanishes within the first nanometer.}
Water structure and orientation at the CaF$_2$–graphene interface (+19.0 mC/m$^2$)
(a) Oxygen density profile ($\rho$) as a function of distance from graphene, highlighting the incipient and first interfacial layers by blue and blue+red shaded areas, respectively. (b) Angular distribution 
P($\theta$) of water molecules in the incipient layer for polarizable (red) and non-polarizable (blue) graphene models.  Vertical dashed lines represent the average value, while horizontal dashed lines represent the uniform distribution present in bulk. (c) 
Same as b) for the whole first interfacial layer, where the impact of graphene polarization becomes negligible.
}
  \label{fig:orientation1D}
\end{figure*}

Figure~\ref{fig:orientation1D} shows the water structure and orientation at the CaF$_2$–graphene interface with a positive surface charge. As observed in the main text for the negatively charged case, the orientation distribution in the incipient layer is affected by the presence or absence of graphene polarization, while these differences vanish when considering the full first hydration layer.

\begin{figure}
\includegraphics[width=1.\textwidth]{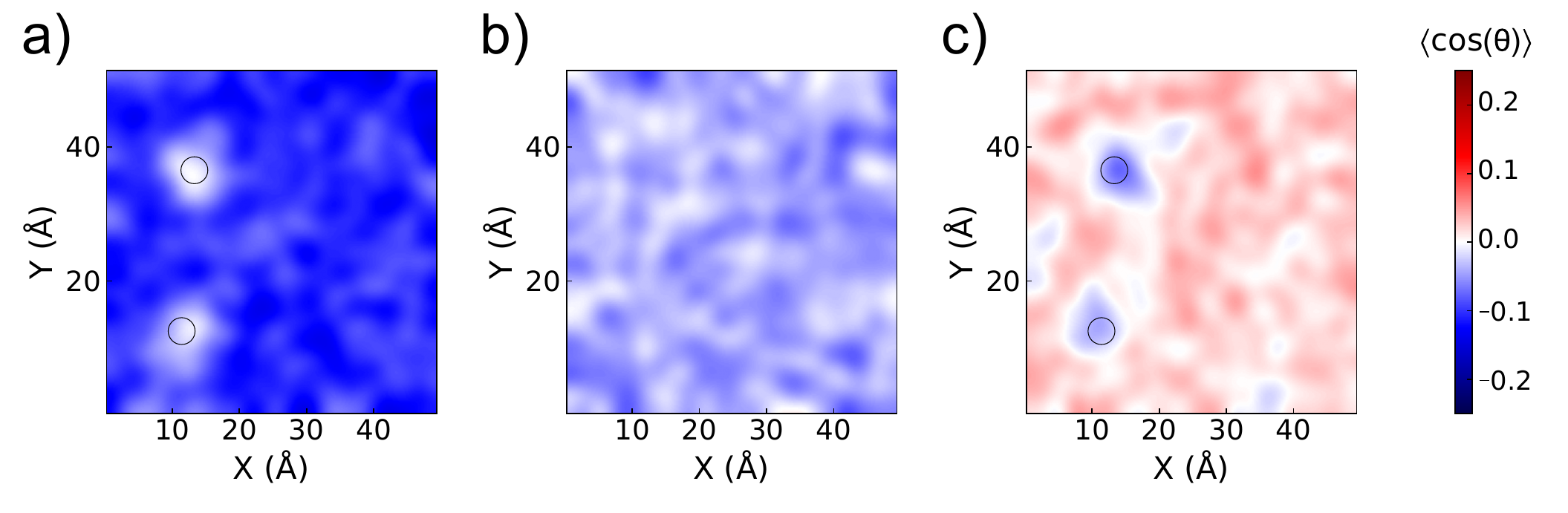}
  \caption{Lateral orientational average of the first interfacial water layer for constant potential simulations with an average (localized) surface charge of 
  a)$\approx$ -10 mC/m$^2$ and  
  b) 0 mC/m$^2$,  and
  c) $\approx$ +10 mC/m$^2$. X and Y are the axes parallel to the surface. Black circles denote the positions of the localized charges beneath the graphene. 
  Unlike the simulations presented in the main text, these supporting simulations were run for five ns after thermalization
  }
  \label{fig:angle-2dmaps}
\end{figure}

Figure~\ref{fig:angle-2dmaps} shows the laterally averaged orientational distribution of water molecules in the first interfacial layer for different localized charge densities.
Consistent with the results discussed in the main text, these profiles reveal that for both positive and negative surface charges, graphene polarization leads to an inverted orientation of interfacial water molecules.

\clearpage
\FloatBarrier

%% file: abstract.tex
\begin{abstract}

Graphene supported on a substrate in contact with water underpins a wide range of processes and technologies, yet its wettability remains controversial. Understanding how substrate charges and graphene's properties influence water organization is crucial.
Here, we combine heterodyne-detected sum-frequency generation (HD-SFG) spectroscopy with molecular dynamics simulations to investigate CaF$_2$-supported graphene interfaces in contact with water. We find that interfacial water orientation is primarily governed by the CaF$_2$ substrate's pH-dependent local electrostatics, confirming graphene's macroscopic wetting transparency. However, at the nanoscale, graphene's polarizability induces a local inversion of water's molecular orientation above substrate charges, revealing subtle structural ordering that is masked in spatially averaged measurements. These insights elucidate the molecular origins of graphene's wetting behavior and suggest new avenues to tailor interfacial phenomena in graphene-based nanofluidic, sensing, and energy applications.

\end{abstract}

%% file: body.tex
\section{Introduction}

The graphene/water interface is central to a wide range of technological applications, including water desalination\cite{Surwade_NatNano_2015}, energy storage and conversion\cite{raccichini_role_2015}, chemo- and biosensing\cite{pumera_graphene_2010}, electrocatalysis\cite{Machado_RSC_2012}, and, more recently, neuromorphic computing\cite{wang_memristor_pnas}. 
In most of these systems, graphene, which is in direct contact with water, is supported on a solid substrate.
At solid/water interfaces, the structure of interfacial water—defined by its orientation and hydrogen-bonding network—dictates microscopic hydrophilicity or hydrophobicity, affects ion adsorption structure and dynamics, mediates surface charging, and modulates interfacial chemical reactivity.\cite{Cyran_PNAS_2019,gonella_water_2021,li_situ_2019,ledezma-yanez_interfacial_2017} Thus, obtaining molecular-level insight into interfacial water at substrate-supported graphene is essential for understanding the mechanisms governing these graphene/water systems.

While both experimental and theoretical studies indicate that suspended, substrate-free graphene on water surface is intrinsically hydrophobic, dominated by dispersive interactions, with an interfacial water structure closely resembling that at the air/water interface\cite{Ohto_PCCP_2018,zhang_water_2020,Wang_Angewandte_2024,Xu_Nature_2023,Rashmi_ACSNano_2025,wang_spectral_2025,yang_nature_2022}—macroscopic contact-angle measurements show that substrate-supported graphene can appear strongly hydrophilic, with its wettability largely dictated by the underlying substrate\cite{taherian_what_2013,KIM_Chem_2021,belyaeva_wettability_2020, kim_wettability_2022}. For this reason, graphene has been described as “wetting transparent”\cite{Rafiee_NatMat_2012}, although this effect is predicted to break down on superhydrophobic or superhydrophilic substrates by theoretical models that consider only dispersive interactions and treat graphene as a purely dielectric sheet.\cite{shih_breakdown_2012}  In practice, however, real substrates generally contain defects or charges, such that the molecular structure of interfacial water on graphene is expected to be strongly affected by substrate electrostatics.\cite{achtyl_aqueous_2015,Franz_free_energy_2013} Indeed, sum-frequency generation (SFG) spectroscopy experiments 
have shown that the aqueous proton–mediated surface chemistry of the substrate governs the interfacial water arrangement through primarily electrostatic interactions\cite{Dreier_JPCC_2019,Montenegro_Nat_2021,Wang_Angewandte_2023,KIM_Chem_2021,Wang_Angewandte_2023,wang_substrate_2023}, with the presence of a graphene monolayer only minimally affecting this surface chemistry.\cite{achtyl_aqueous_2015,Wang_Angewandte_2023} These findings collectively point to electrostatic interactions—rather than dispersive interactions—as the dominant origin of graphene’s apparent “wetting transparency” in oxides and ionic substrates\cite{KIM_Chem_2021,Wang_Angewandte_2023,Wang_Angewandte_2023,wang_substrate_2023,Montenegro_Nat_2021}

Graphene’s commonly observed transparency to substrate electrostatic interactions is frequently ascribed to its atomic-scale thickness.\cite{driskill_wetting_2014,Carlson_JCPL_2024} However, due to its pronounced polarizability \cite{Ojaghlou_ACSNano_2020}, a key unresolved question is whether substrate-generated electric fields penetrate graphene to affect interfacial water directly, or whether the fields first induce polarization within graphene itself, which then mediates the interactions with water across the layer. Disentangling this mechanism is essential for a fundamental understanding of how graphene’s interfacial properties originate at the molecular level.

To address this question, we combine heterodyne-detected sum-frequency generation (HD-SFG) spectroscopy, which experimentally probes the molecular structure of water at CaF$_2$/water interfaces both with and without a graphene monolayer, with molecular dynamics simulations to provide a nanoscale view of how substrate electrostatics influence water across graphene. Comparing the interfacial water structures at the two interfaces, therefore, enables us to access the molecular origin of graphene’s wetting transparency experimentally. Our spectra reveal that the interfacial water structures at the two interfaces are similar, although both vary strongly with pH due to aqueous proton–mediated substrate chemistry, demonstrating that substrate electrostatics govern interfacial water arrangement and constitute the molecular basis of the observed wetting transparency. 
At the same time, our atomistic simulations show that graphene’s polarizability actively responds to electric fields originating from the substrate, significantly modifying the interfacial water structure compared to a (hypothetical) non-polarizable graphene. Yet, when averaged over the first interfacial water layer, the well-known transparency of graphene emerges as the local effects are washed out. This combined approach provides a molecular-level picture of the interplay between substrate charges, graphene, and interfacial water. Our findings have important implications for designing and interpreting graphene-based systems in aqueous environments.

\subsection{Microscopic Interfacial Water Structure}

\begin{figure*}
\includegraphics[width=.99\textwidth]{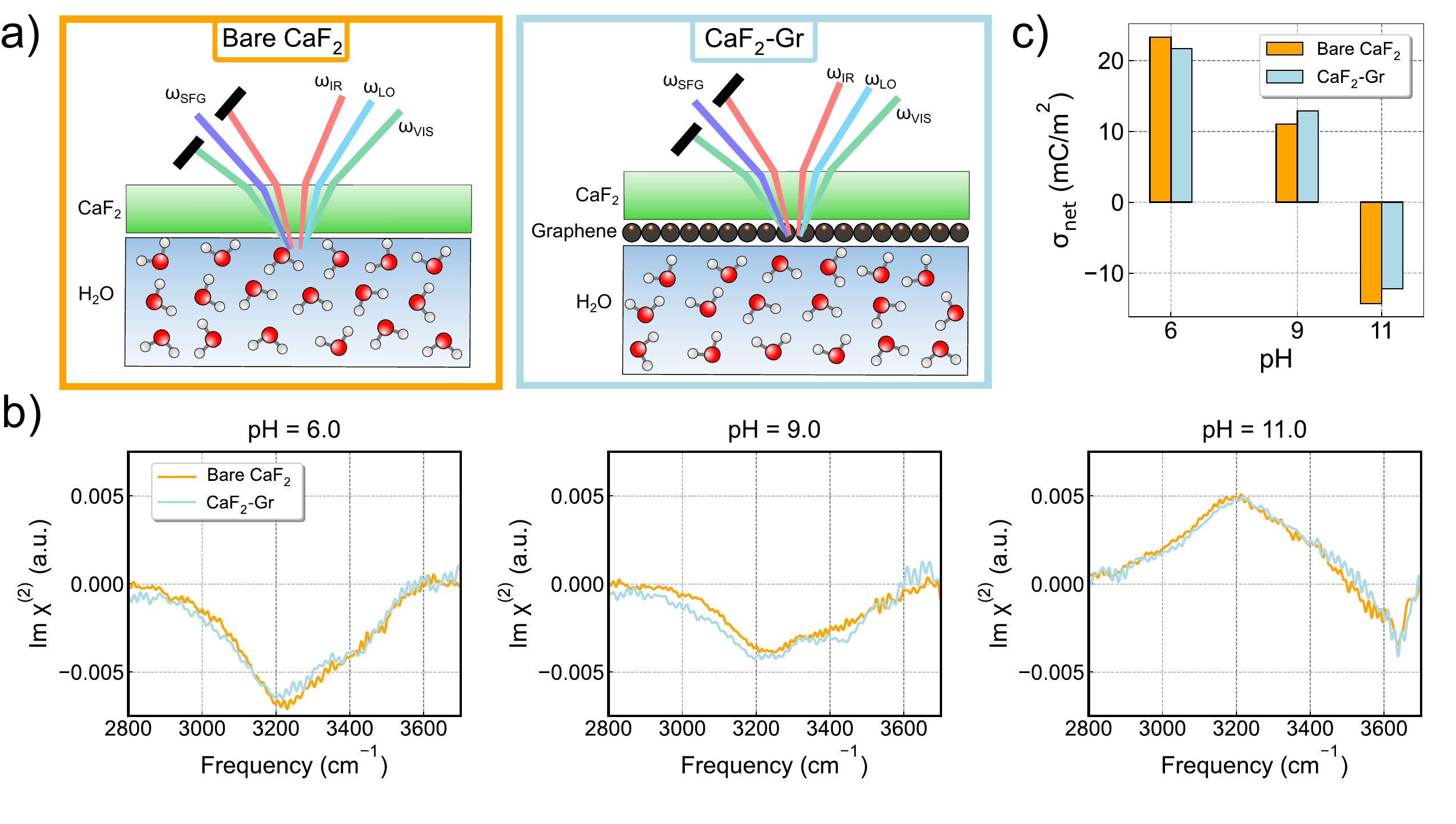}
  \caption{\textbf{Interfacial water structure at CaF$_2$/water and CaF$_2$–Gr/water interfaces under varying pH conditions is dominated by electrostatic effects and remains largely unaffected by graphene at the macroscopic scale.} a) Schematic representation of the charged water interfaces. b) The O-H stretching Im$\chi^{(2)}$ spectra were measured at varying pH conditions. c) Net surface charge density inferred from the bulk contributions by comparing SFG signals at different NaCl concentrations, at varying pH conditions for the two interfaces.
  \label{fig:SFG}}
\end{figure*}

HD-SFG is a surface-specific vibrational technique that gives access to the complex $\chi^{(2)}$ spectrum, whose imaginary component (Im$\chi^{(2)}$) directly reports on interfacial water orientation and hydrogen-bonding\cite{Nihonyanagi_JCP_2009,Urashima_JCPL_2018,shen_phase-sensitive_2013,Richmond_CCl4_science}. These structural features are highly sensitive to surface forces, including electrostatic and dispersive interactions\cite{wen_unveiling_2016,Wang_Angewandte_2024,wang_spontaneous_2025,Shen_ChemRev_2006,Richmond_CCl4_science}. In the following, we consider the bare CaF$_2$/water and CaF$_2$-supported graphene (CaF$_2$–Gr)/water interfaces and report the effective second-order nonlinear susceptibility, $\chi^{(2)}$, in the O–H stretching region (2900-3700 cm$^{-1}$) using HD-SFG spectroscopy. Experiments were performed at varying solution pH levels with a fixed ionic strength of 10 mM NaCl in a custom-built flow cell; the low salt concentration was chosen to maintain ionic strength while minimizing specific ion effects. The HD-SFG experiments were conducted at the SSP polarization combination, where the letters denote the polarizations of the SFG, visible, and infrared beams, respectively (see Methods). A schematic of the sample composition and beam geometry of the HD-SFG measurement is shown in Fig. \ref{fig:SFG}a. Additional information on sample preparation and HD-SFG measurements is provided in Refs.\cite{Montenegro_Nat_2021,Wang_Angewandte_2023,Wang_Nature_2023,Dreier_JPCC_2019}, and recalled in Methods.
  
The pH-dependent data are shown in Fig. \ref{fig:SFG}b. At neutral pH ($\sim$6), the Im$\chi^{(2)}$ spectra at the bare CaF$_2$/water interface predominantly display a broad negative peak from 3000 to 3600 cm$^{-1}$, originating from hydrogen-bonded (H-bonded) O–H groups of interfacial water. The negative sign indicates that, on average, these O–H groups are oriented with their dipoles pointing down toward the bulk solution. This alignment of  interfacial water arises from the positively charged CaF$_2$ surface at neutral pH, given that its isoelectric point lies around pH = 9--10.~\cite{Assemi_Langmuir_2006,CaF2_pzc_1997} In fact, at such a charged interface and under the employed ionic strength (10 mM), the SFG response contains not only the surface contribution from the topmost 1–2 molecular layers, but also a dominant bulk contribution arising from the penetration of the electrostatic field into the solution, which induces alignment and polarization of water molecules in the diffuse layer\cite{wen_unveiling_2016,ohno_second-order_2017,reddy_bulk_2018,Fellows_EDL_JPCL_2024,Wen_Momemtum_SA_2023}. This bulk contribution thus serves as a sensitive probe of the interfacial electrostatics. 

We further estimate the net surface charge density from the bulk contributions by comparing SFG signals measured at different ionic strengths, following previously established protocols.\cite{wen_unveiling_2016,seki_real-time_2021,Wang_Nature_2023} The resulting net surface charge density is $\sim$23 mC/m$^2$ (Fig. \ref{fig:SFG}c), consistent with the observed interfacial water alignment. These observations are consistent with previous studies.\cite{Becraft_Langmuir_2001, khatib_water_2016} 

The Im$\chi^{(2)}$ spectrum at the CaF$_2$–Gr/water interface under neutral pH closely resembles that at the bare CaF$_2$/water interface (Fig.~\ref{fig:SFG}b), with the corresponding surface charges also in good agreement, within experimental uncertainty (Fig.~\ref{fig:SFG}c). This suggests that the organization of interfacial water at the CaF$_2$–Gr/water interface is dominated by the CaF$_2$’s electrostatic interactions with water. Several recent studies have reported alkane (or hydrocarbon) adsorption from water onto graphite surfaces~\cite{bonagiri_2025_graphite,arvelo_2022_interfacial}, and other reports suggest that water suppresses such adsorption.\cite{Li_acsnano_2016_hydrocarbon}. Thus, verifying surface cleanliness is therefore essential, as even trace levels of hydrocarbon adsorption have been shown to alter graphene’s interfacial properties and compromise its apparent transparency.
\cite{Mucksch_JPCC_2015,Carlson_JCPL_2024,Li_NatMat_2013}
Notably, the absence of C–H vibrational features (2850–2950 cm$^{-1}$) in all Im$\chi^{(2)}$ spectra confirms that our samples are free from noticeable hydrocarbon contamination, underscoring their cleanliness.

To further confirm the electrostatic dominance of CaF$_2$, we measured the Im$\chi^{(2)}$ spectra at both interfaces while varying pH to tune the surface charge density of CaF$_2$, which is set by interfacial acid–base equilibria.~\cite{Assemi_Langmuir_2006} Because protons are transmitted through graphene\cite{hu_proton_2014,achtyl_aqueous_2015}, the graphene monolayer minimally perturbs these equilibria,\cite{achtyl_aqueous_2015} and the interfacial water structure is expected to respond similarly to pH changes at both interfaces, as observed in Fig.~1b. At both interfaces, increasing pH causes the H-bonded O–H band to diminish and reverse sign, becoming positive above pH=10. Simultaneously, a negative peak emerges at 3630~cm$^{-1}$, corresponding to the Ca–O–H stretch of hydroxylated CaF$_2$.~\cite{khatib_water_2016, Abdelghani_NatMat_2025,Wang_Angewandte_2023,Montenegro_Nat_2021}. These Ca–O–H groups originate from adsorbed hydroxyl ions, which can form between the CaF$_2$ substrate and graphene, as some water molecules are trapped between the two \cite{wang_chemistry_2023,Wang_Angewandte_2023,montenegro_asymmetric_2021}. Proton equilibration across the graphene sheet generates hydroxyl groups at the calcium fluoride surface. These spectral changes reflect the reorientation of interfacial water in response to the CaF$_2$ surface charge, which switches from positive ($\sim$23 mC/m$^2$) to negative ($\sim-14$ mC/m$^2$, Fig.~1c), confirming that water organization at the CaF$_2$–Gr/water interface is dominated by CaF$_2$ electrostatics. We also conducted measurements on the SiO$_2$-supported graphene/water interface, and the results consistently indicate the dominant role of the substrate’s electrostatics (Fig. S1 in the SI). These results are fully consistent with the well-documented apparent wetting transparency of graphene, as inferred from macroscopic contact-angle measurements\cite{taherian_what_2013,KIM_Chem_2021,belyaeva_wettability_2020, kim_wettability_2022,Rafiee_NatMat_2012}, and further show that this transparency holds even at the microscopically averaged structure of interfacial water.

An alternative explanation for the observed transparency involves partial cancellation of VSFG signals arising from water molecules located on opposite sides of graphene supported on charge-neutral substrates.\cite{Ohto_PCCP_2018,Dreier_JPCC_2019,hou_2025_waters_stackedgraphene,Ojaghlou_ACSNano_2020}
While such cancellation cannot be entirely ruled out in our experiments, it is unlikely to serve as the primary origin, as it would not be expected to occur with the same magnitude under different pH and surface-charge conditions. The nanoconfined water between graphene and the substrate spans only a few molecular layers\cite{wang_interfaces_2025,Lee_Jacs_2014}, whereas in a bulk 10 mM NaCl solution, the SFG response primarily arises from the polarization and reorientation of water molecules within the diffuse layer, which extends over approximately the Debye length ($\sim$3.4 nm). Therefore, if present, this effect is expected to play only a minor role.

\subsection{Nanoscopic Interfacial Water Structure}

HD-SFG probes an interfacial region with a lateral dimension of $\sim$100--200~$\mu$m, whereas CaF$_2$ substrate charges are localized at a density of roughly one per few square nanometers ($\sim$3--10 nm$^2$) at neutral pH. To further unveil how substrate electrostatics influence water across graphene at the nanoscale, we performed molecular dynamics simulations of the CaF$_2$-supported graphene/water interface. 

We carried out classical molecular dynamics simulations of 1 M NaCl aqueous solutions at CaF$_2$-supported graphene surface. 
The simulation setup consisted of a liquid slab confined between two solid surfaces, each formed by a single layer of graphene supported on CaF$_2$. Localized surface charges were introduced by modifying the charge of selected atoms in the topmost CaF$_2$ layer, while overall charge neutrality was preserved by adding compensating Na$^+$ or Cl$^-$ ions to the aqueous phase.
Graphene polarization was described using the Siepmann–Sprik constant-potential method,\cite{siepmann_influence_1995}, in which Gaussian charge distributions centered on carbon atoms are adjusted at each timestep to minimize the electrostatic energy. To isolate the effect of graphene polarization, we additionally performed reference simulations with non-polarizable graphene, achieved by setting the carbon atom charges to zero. Further computational details are provided in the Methods section.

\begin{figure*}
\includegraphics[width=.45\columnwidth]{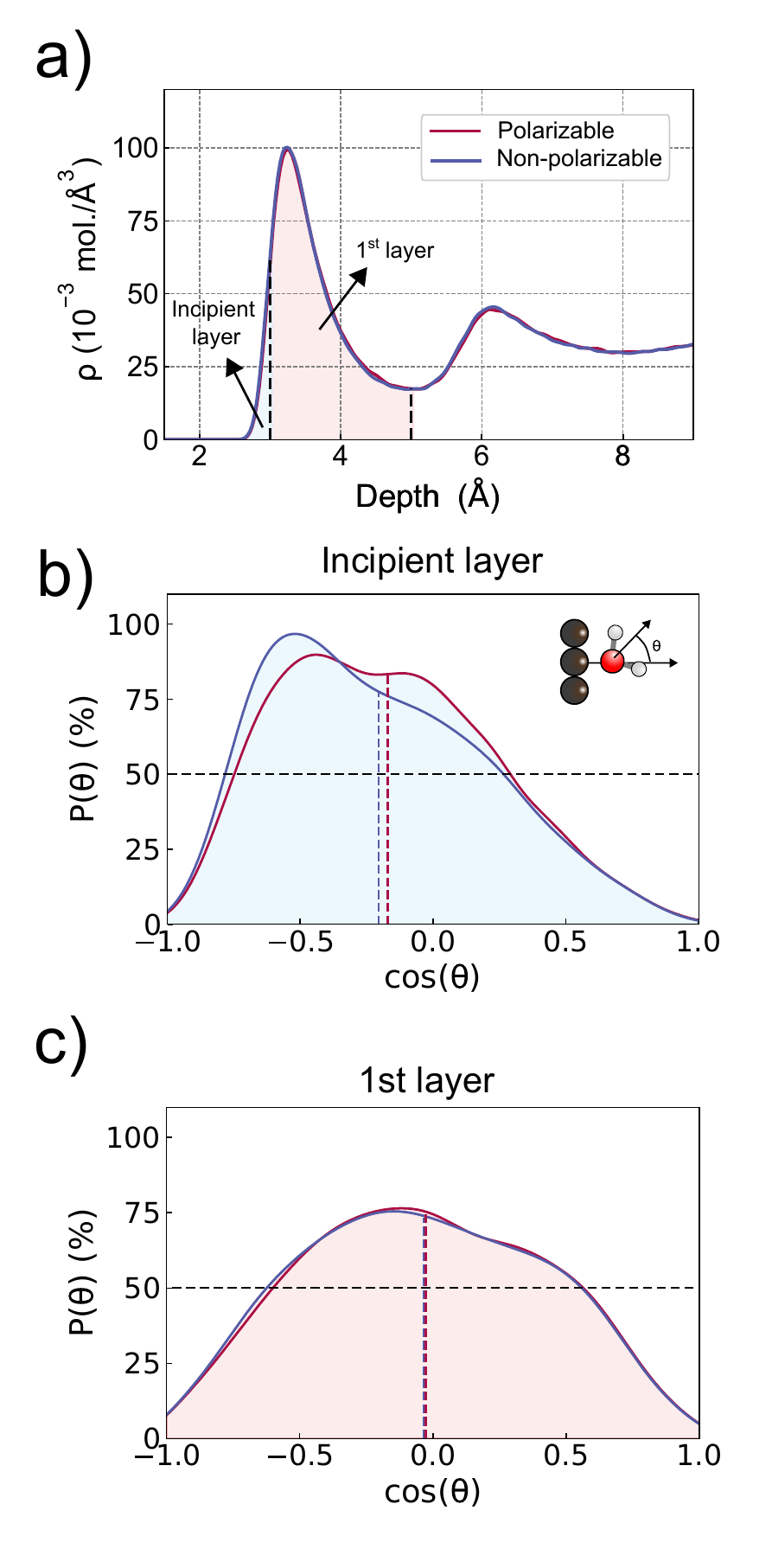}
  \caption{\textbf{Graphene-induced polarization reshapes the initial interfacial water layer but vanishes within the first nanometer.}
Water structure and orientation at the CaF$_2$–graphene interface ($-19.0$ mC/m$^2$)
(a) Oxygen density profile ($\rho$) as a function of distance from graphene, highlighting the incipient and first interfacial layers by blue and blue+red shaded areas, respectively. The incipient layer is defined as the region within 3~\AA\, of the graphene (see main text). (b) Angular distribution 
$P(\theta$) of water molecules in the incipient layer for polarizable (red) and non-polarizable (blue) graphene models.  Vertical dashed lines represent the average value, while horizontal dashed lines represent the uniform distribution present in bulk. (c) 
Same as b) for the whole first interfacial layer, where the impact of graphene polarization becomes negligible.
}
  \label{fig:orientation1D}
\end{figure*}

Figure \ref{fig:orientation1D}a shows the water density profiles from polarizable and non-polarizable simulations with a surface charge of $-19.0$~mC/m$^2$, close to the experimental value at pH=11. The two profiles are nearly identical, indicating that graphene polarization has little effect on the overall distribution of water molecules\cite{Ojaghlou_ACSNano_2020}. In both cases, pronounced layering extends up to approximately 1 nm from the surface, consistent with previous simulations based on classical force fields~\cite{Zhan_JPCL_2020}, density functional theory~\cite{Tocci_NanoLett_2014}, and machine-learning potentials~\cite{Gading_PNAS_2024}.
Figure \ref{fig:orientation1D}b depicts the distribution of orientations, given by the angle between the water dipole and the surface normal (see inset), for the incipient layer, defined as molecules located within 3~\AA{} of the graphene surface (panel a). Non-polarizable simulations yield a more structured distribution with a maximum at $\cos(\theta)\approx -0.6$ (where $\theta$ is the angle between the water dipole and the surface normal), whereas graphene's polarization reduces this ordering. The vertical lines indicate the mean orientation, which remains similar in both cases.
Fig. \ref{fig:orientation1D}c shows the same analysis for the first complete hydration layer (also defined in panel a). In this case,  the distributions overlap almost perfectly, showing that the differences observed in the incipient layer decay within the first half nanometer. While the precise cutoff is somewhat arbitrary, the conclusion that the orientational distribution varies strongly across the first water layer is robust. Moreover, 
 estimating the relative contribution of each region to the VSFG signal by approximating it as the integral of $\rho(z) \times \langle \cos(z) \rangle$ \cite{pop_model_2024}, we find that the incipient layer contributes approximately 5\% to the total signal. This indicates that such subtle interfacial effects are difficult to capture experimentally using VSFG.  In the SI, we show results for an opposite surface charge (+19.0 mC/m$^2$) and find comparable behavior.  

\begin{figure*}
\includegraphics[width=0.99\columnwidth]{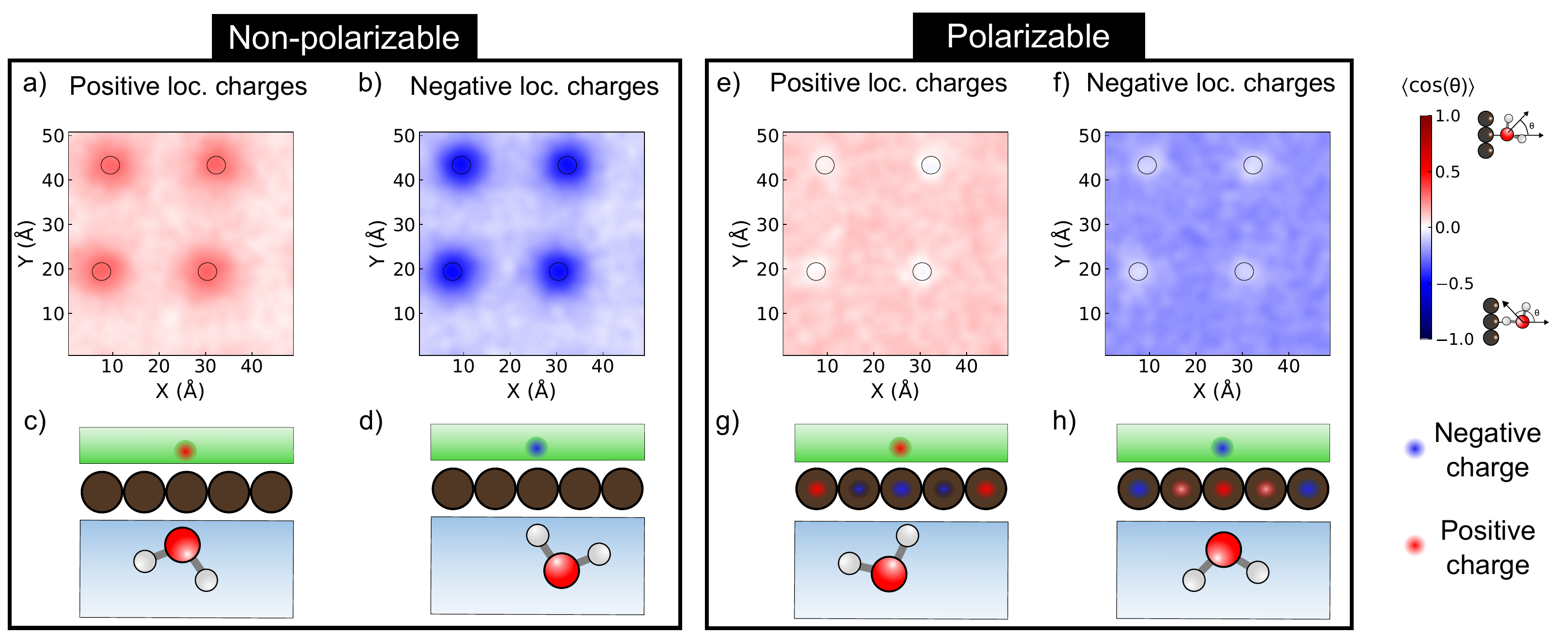}
  \caption{\textbf{Graphene polarizability inverts the effective sign of the localized charge, inducing an inversion in the orientation of water molecules}. a) Spatial orientational average of the incipient interfacial water layer for non-polarizable graphene with a surface charge of $\approx$ +20 mC/m$^2$. 
b) Same as a), for a surface charge of $\approx -20$ mC/m$^2$.
Black circles denote the positions of the localized charges. 
c-d) show cartoon representations of the interface, the charges, and water orientations corresponding to a-b),  respectively. 
Blue and red dots represent negative and positive charges, respectively. 
Panel e-h) show the corresponding results for polarizable graphene. 
\label{fig:orientation2D}
  }
\end{figure*}

The one-dimensional probability distributions shown in Fig. \ref{fig:orientation1D} capture the average orientational behavior of interfacial water but hide the spatial heterogeneities present in the system. To resolve these effects, we further analyze the spatial distribution of water molecules from the simulations. Figures \ref{fig:orientation2D}a–b display the two-dimensional orientation probability maps of the incipient interfacial layer for positive and negative localized surface charges in the presence of non-polarizable graphene. As expected, the localized charges induce a heterogeneous orientational pattern.
In this case, the graphene is electrostatically transparent, and the water molecules behave as expected, with hydrogen atoms pointing
away from the localized positive charges and toward the localized negative charges. This is illustrated in graphical depictions in panels c and d, respectively. 

Figures \ref{fig:orientation2D}e–f present the corresponding two-dimensional orientation probability for the case of polarizable graphene.
Surprisingly, the observed orientation is opposite to the intuitive expectation: instead of hydrogen atoms pointing away from (toward) the positive (negative) charge, the simulations show that they orient toward (away from)  it (see panels g–h for graphical depictions). This counterintuitive effect arises from the polarizability of the graphene sheet. A localized positive (negative) charge on the substrate induces a corresponding negative(positive) image charge on the graphene directly above it. As a result, water molecules situated above the localized substrate charge align toward (or away from) it.
At larger distances, water molecules exhibit the opposite orientation, which is more pronounced than in systems with homogeneous surface charges (see SI). This compensatory effect, rooted in charge conservation within the graphene, yields a similar net orientation when averaged over extended regions. In the SI, we show that this behavior is robust across a range of charge densities. 
Similar polarization-driven mechanisms have been proposed to explain the reduction of contact ion pairing at water/graphene interfaces,\cite{Fong_NanoLett_2024} the mild stabilization of hydroxide at the interface,\cite{Advincula_ACSNano_2025} and the attraction of like charges across graphene sheets.\cite{Ojaghlou_ACSNano_2020}

These results highlight the fundamentally non-transparent behavior of a single graphene monolayer at the nanoscale when its polarization is properly accounted for. While graphene appears transparent to the underlying substrate at the macroscopic scale, its polarization enables modulation of the nanoscopic structure of the incipient interfacial water. This reveals the potential mechanism that could influence key interfacial phenomena, such as ionic conductivity in nanopores and water flow.\cite{mangaud_chemisorbed_2022}
Let us take the example of the solid-liquid friction. 
For uniformly charged surfaces, the friction coefficient is known to increase with surface charge density~\cite{Jing_JCIS_2015,Joly_JCP_2006}.
Moreover, it has been shown by Joly \textit{et al.}, that the spatial distribution of localized charges on graphene has a significant impact on interfacial friction~\cite{Xie_PRL_2020}.
In their study, the enhanced friction was attributed to the formation of bound ions, which was particularly pronounced in the model employed, where fixed localized charges were placed directly on the carbon atoms.
While we expect that this mechanism is less prominent when localized charges are positioned beneath the graphene, and polarization effects are taken into account, we expect that the increased surface roughness may instead become the dominant factor influencing the friction coefficient~\cite{Tocci_NanoLett_2014}.
In addition, strong local electric fields near carbon atoms alter graphene’s electronic properties~\cite{Castro_RevModPhys_2009}, potentially influencing the electronic contribution to the friction~\cite{Greenwood_natCom_2023,Bui_NanoLet_2023,Kavokine_Nature_2022}.
Future work will focus on quantifying and characterizing these dynamical effects in experimentally realizable setups.

\section{Conclusions}

In summary, we have presented a combined experimental and theoretical study on the apparent transparency of graphene.  Graphene is often assumed to be partially or fully transparent to its underlying substrate~\cite{Rafiee_NatMat_2012,KIM_Chem_2021,Ghoshal_Langmuir_2019}, yet its molecular origin remains elusive or controversial.
Our experimental results indicate that substrate electrostatics indeed govern interfacial water arrangement, providing the molecular basis for the observed apparent wetting transparency of graphene in oxides and ionic substrates. Simulations support this conclusion; however, at the nanoscale, they reveal that graphene does not behave as a truly transparent layer. While the macroscopic water structure appears to be unaffected by the underlying charge distribution, the microscopic response reveals significant influence from it.
In particular, we show that the polarizability of graphene gives rise to a localized induced charge that acts as a mirror of the substrate’s charge distribution. This mirrored charge alters the local electric field, thereby modifying the orientation of interfacial water molecules.
We suggest that this effect could be directly confirmed using nanoscale chemical imaging techniques at solid–liquid interfaces, such as tip-enhanced Raman spectroscopy~\cite{Kumar_Nanoscale_2018,Litman_JPCL_2023}.
Beyond advancing our fundamental understanding of water/graphene interfaces, these insights may be relevant for the fabrication of nanoscale devices based on graphene or other 2D materials. They  suggest new possibilities to influence coupling between materials (solids or liquids) positioned on opposite sides of the 2D layer~\cite{Avetisyan_PRB_2009} and control the spatial dependence of water flow by, for example, tuning the classical and electronic/quantum friction across the interfaces~\cite{Greenwood_natCom_2023,Bridge_JCP_2024,Bui_NanoLet_2023,Coquinot_NatNano_2025}.

\section{Methods}

\subsection{Molecular dynamics simulations}

Molecular dynamics simulations were performed using the LAMMPS code~\cite{thompson_lammps_2022}. The system consisted of pure water confined between two solid interfaces, each composed of a single graphene layer and three layers of CaF$_2$. 

Graphene sheets were constructed with 960 carbon atoms, a C-C bond length of 1.426~\AA. We employed a cubic simulation box of dimensions 49.400~\AA~$\times$ 51.336~\AA~$\times$ 85.000~\AA~ with periodic boundary conditions in all three directions. 
The graphene sheets were placed 40~\AA~apart, and each system contained 3000 water molecules confined between them. For computational efficiency, the positions of the substrate and graphene atoms were kept fixed. The simulation box dimensions were chosen to be large enough to prevent interactions between the two opposing interfaces. In the system without localized charges, 56 Na$^+$ and Cl$^-$ ions were included to achieve an approximately 1~M concentration. 
Localized charges were introduced by modifying the charge of selected atoms in the topmost CaF$_2$ layer. Charge neutrality was maintained by adding compensating Na$^+$ or Cl$^-$ ions in the aqueous region. Graphene polarization was modeled using the Siepmann-Sprik constant-potential method~\cite{siepmann_influence_1995} as implemented in the Electrode package of LAMMPS~\cite {ahrens-iwers_electrode_2022}, with a reciprocal Gaussian width of 1.979~\AA$^{-1}$. The simulation with polarizable graphene was performed by setting the potential difference between the two sheets to 0 V, while allowing the atomic charges of carbon to fluctuate. While not explicitly enforced, we verified that the net charge on the graphene electrodes remains constant and close to 0. %
Within this framework, graphene is treated as a conductor and is thus expected to capture the dominant polarization and induction effects relevant to its behavior in polar environments\cite{ZHONG201720}.

Water molecules were modeled using the SPC/E model\cite{Berendsen_JCP_1987}. Ionic interactions were described using electronic continuum correction rescaled force fields developed by the Jungwirth group~\cite{Kohagen_JPCB_2016, Pluharova_MolPhys_2014}. Lennard-Jones parameters for carbon atoms were taken from Ref.~\cite{Rafiee_NatMat_2012}. All carbon, calcium, and fluoride atoms were held fixed at their lattice positions. All simulations were carried out in the NVT ensemble at 300~K using a Nosé-Hoover thermostat with a damping parameter of 200~fs. The equations of motion were integrated using the velocity-Verlet algorithm with a time step of 1.0~fs. Bond lengths involving hydrogen atoms were constrained using the SHAKE algorithm. Initial configurations were generated with Packmol \cite{Martínez_JCC_2009} and thermalized for 1~ns. Unless specified otherwise, production runs for the simulations, 
from which all reported averages were computed, lasted 25~ns.

\subsection{Sample preparation and Vibrational SFG}

The CaF$_2$-supported graphene samples were prepared using a polymer-assisted wet transfer method with commercial CVD-grown graphene, following procedures detailed in previous studies.\cite{Montenegro_Nat_2021,Wang_Nature_2023,Wang_Angewandte_2023} Graphene was purchased from Graphenea, Inc., and CaF$_2$ substrates were obtained from Korth Crystals GmbH. Comparable results were also obtained using graphene from Grolltex, Inc. and CaF$_2$ substrates from Shalom EO and PI-KEM Ltd. Following assembly in the custom-built Teflon flow cell, the cell was first filled with pH 3 water for 10 minutes to refresh the CaF$_2$ surface and then rinsed three times with neutral water using a syringe pump. To minimize flow-induced effects, the sample was allowed to equilibrate for 20 minutes after each solution exchange before SFG data acquisition.

All the complex $\chi^{(2)}$ spectra were acquired using an HD-SFG spectrometer in a non-collinear beam geometry, powered by a Ti:Sapphire regenerative amplifier laser system operating at a central wavelength of 800 nm, with a pulse width of ~40 fs, pulse energy of 5 mJ, and a repetition rate of 1 kHz, as detailed in refs. \cite{Wang_Nature_2023,seki_real-time_2021}. The incident angles for the IR, visible, and LO beams (in CaF$_2$) were 33°, 39°, and 37.6°, respectively. The measurements were performed in a dried air atmosphere to prevent spectral distortion from water vapor.

The phase measurement method followed previously established procedures \cite{Wang_Nature_2023,Wei_JACS_2023}. Briefly, a ~100 nm-thick gold layer was coated on half of the CaF$_2$-supported graphene sample and used as a 
intermediate phase reference. The phase of the gold-coated reference was determined by measuring the CaF$_2$-supported graphene/D$_2$O signal at the same position used for the CaF$_2$-supported graphene/water measurements, under the assumption that the response of the CaF$_2$-supported graphene/D$_2$O interface is purely real~\cite{Yamaguchi_JCP_2011}.
HD-SFG measurements were performed using a custom-made flow cell that enabled solution exchange, as described in detail in refs. \cite{Wang_Nature_2023,Wang_Angewandte_2023}. 
A height displacement sensor (CL-3000, Keyence) was used to check the sample height.